\documentclass{article}
\usepackage{amssymb}
\usepackage{german} 
\usepackage{graphicx} 
\usepackage{utopia}

\begin{document}
\tolerance=10000
\def\shalf{\hbox{${\textstyle{\frac{1}{2}}}$}}

\title{Das R"atsel der kosmischen Vakuumenergiedichte und die \\ 
       beschleunigte Expansion des Universums}

\author{Domenico Giulini und  Norbert Straumann \\
        Institut f"ur Theoretische Physik       \\
        der Universit"at Z"urich                \\
        Winterthurerstrasse 190                 \\
        CH-8057 Z"urich, Schweiz                  }
  
%\thanks{}
\maketitle

\begin{abstract}
Die Grundprinzipien der Allgemeinen Relativit"atstheorie lassen 
die Freiheit, die Feldgleichungen zus"atzlich mit dem sogenannten
kosmologischen Term auszustatten, der m"oglicherweise ganz oder 
teilweise durch die Quantenfluktuationen der Materiefelder 
interpretiert werden kann. Je nach Vorzeichen und Gr"o"se kann dieser 
in gewissen Entwicklungsphasen des Universums zu einer beschleunigten
oder verz"ogerten Expansion f"uhren. Neueste kosmologische
Beobachtungen scheinen daf"ur zu sprechen, dass wir uns 
gegenw"artig in einer Phase \emph{beschleunigter} Expansion 
befinden. Wir geben zun"achst einen Abriss der Geschichte und
fundamentalen Problematik des kosmologischen Terms und stellen 
dann die gegenw"artigen Evidenzen f"ur einen positiven Wert vor, 
der eben zu einer beschleunigten Expansion f"uhrt.
\end{abstract}

\section*{1 Einleitung}
Astronomische Beobachtungen der letzten zwei Jahre deuten  
mit zunehmender Deutlichkeit auf eine \emph{beschleunigte}
Expansion des Universums. Dies klingt zun"achst sehr 
erstaunlich, sind wir doch daran gew"ohnt, dass die
Gravitationskraft immer anziehend ist und deshalb auch die 
gro"sr"aumige Fluchtbewegung der Galaxien abbremsen sollte.
Tats"achlich l"a"st sich eine kosmische Absto"sung weder 
bei den Planetenbahnen in unserem Sonnensystem noch in den 
Bewegungen von Galaxiengruppen beobachten.

Die  M"oglichkeit, eine kosmische Absto"sung in den
Feldgleichungen der Allgemeinen Relativit"atstheorie zu
implementieren, erkannte Einstein schon Anfang 1917, etwas
mehr als ein Jahr nach Aufstellung dieser Gleichungen, als er 
versuchte, diese auf das ganze Universum anzuwenden~\cite{Einstein-1917}. 
Dabei bemerkte er, dass trotz der stark einschr"ankenden
formalen Prinzipien es noch gestattet ist, die urspr"ungliche 
Form der Feldgleichungen um einen sehr einfachen Zusatzterm 
zu erweitern, den man seither den kosmologischen Term nennt. 
Dieser beinhaltet einen noch unbestimmten multiplikativen
Parameter, den man allgemein mit $\Lambda$ bezeichnet und 
f"ur den sich die Bezeichnung \emph{kosmologische Konstante}
eingeb"urgert hat.
F"ur positives $\Lambda$ wirkt der kosmologische Term wie
eine konstante, den Raum auseinanderziehende Spannungskraft.
Da sich diese im Raum additiv verh"alt, kann sie unter Umst"anden auf 
gro"sen Skalen "uber eine gravitative Abbremsung dominieren.

Sp"ater hat Einstein seine Einf"uhrung des kosmologischen 
Terms bedauert, aus heutiger Sicht wahrscheinlich zu Unrecht. 
Zun"achst "uberzeugt sein einfaches Argument nicht, die 
Gleichungen seien ohne kosmologischen Term einfacher. Denn es 
ist eine allgemeine Erfahrungstatsache, dass formale 
Komplikationen, die nicht durch fundamentale Prinzipien verboten 
werden, von der Natur auch sehr h"aufig realisiert werden, z.B.
im Standardmodell der Elementarteilchenphysik. 
Vielmehr ist ein kosmologischer Term sogar zu erwarten, da nach 
heutigen Quantenfeldtheorien auch der nominell leere Raum (das Vakuum)
i.A. nicht verschwindende Energie besitzt (siehe Anhang~2). Die
gravitative Wirkung dieser sogenannten \emph{Vakuumenergie} ist aber 
genau die eines kosmologischen Terms. Obwohl dieser Term im Falle 
einer positiven kosmologischen Konstante einer positiven 
Energiedichte entspricht, f"uhrt er trotzdem zu einer gravitativen
Absto"sung. Dies werden wir noch ausf"uhrlicher diskutieren.

Tats"achlich wundern sich die theoretischen Physiker schon 
lange dar"uber, warum die Vakuumenergiedichte nicht sehr viel
gr"o"ser ist,  als es die astronomischen Beobachtungen zulassen.
Dieser mysteri"ose Aspekt der kosmologischen Konstante  
wird seit geraumer Zeit als eines der tiefsten R"atsel
der Grundlagenphysik angesehen.

F"ur Astronomen und Kosmologen gibt es einen zweiten Aspekt 
im Zusammenhang mit einer m"oglicherweise nicht verschwindenden 
kosmologischen Konstanten. Da n"amlich die Vakuumenergiedichte 
zeitlich konstant ist, aber andererseits die Energiedichte der
Materie mit der kosmischen Expansion abnimmt, ist es 
h"ochst erstaunlich, dass beide ausgerechnet zum
gegenw"artigen Zeitpunkt von vergleichbarer Gr"o"se sind, 
wie die noch zu besprechenden Messungen zeigen.
L"osungsversuche dieses "`kosmischen Koinzidenzproblems"'
werden zur Zeit lebhaft diskutiert. Wir werden am Schluss
dieses Aufsatzes kurz darauf eingehen.
 
\section*{2 Eine kurze Geschichte des kosmologischen Terms}
Die Vorstellung eines dynamisch (expandierenden) Universums
hat sich erst im 20. Jahrhundert durchgesetzt, obwohl schon  
Newton (nach Intervention Bentleys) klar war, dass die 
universell anziehende Gravitation kein statisches Universum 
zul"asst. Der Zusammensturz der Fixsterne 
schien ihm unvermeidlich, wie u.A. aus folgendem Zitat aus einem
seiner ber"uhmten Briefe an Bishop Bentley vom 25.~Februar 1693
hervorgeht~(siehe dazu die Diskussion in \cite{Harrison-1986}, 
wo auch einige der relevanten Textstellen dieser interessanten 
Korrespondenz wiedergegeben sind):

% ANFANG ZITAT NEWTON
``And tho all ye matter were at first divided into several
systems and every system by a divine power constituted like
ours: yet would the outward systems descend towards the
middlemost so yt this frame of things could not always
subsist without a divine power to conserve it''.
%% ENDE ZITAT

Versucht man aber tats"achlich die
Newtonschen Gravitationsgesetze auf das ganze Universum anzuwenden,
so wird man einer ganzen Reihe von Problemen begegnen:
So sind die Ausdr"ucke f"ur das Gravitationspotential und die Kraft
nur dann wohldefiniert, wenn die Massendichte f"ur gro"se r"aumliche 
Abst"ande schneller als $r^{-2}$ bzw. $r^{-3}$ abf"allt. Das unendlich 
gedachte Universum kann also nicht eine im Mittel homogene 
Massenverteilung endlicher Dichte tragen. Gegen die Annahme 
eines mehr oder weniger r"aumlich lokalisierten Universums
im statistischen Gleichgewicht spricht aber der
sogenannte "`Ver"odungseinwand"', nach dem im Laufe der Zeit immer
wieder Sterne ins r"aumlich Unendliche  "`abdampfen"' k"onnen.
Diesem k"onnte man zwar mit der Annahme eines im r"aumlich Unendlichen
nicht konstanten, sondern ansteigenden Gravitationspotentials begegnen,
doch spricht gegen dieses die Beobachtungstatsache kleiner lokaler
Relativgeschwindigkeiten der Sterne.

\subsection*{2.1 Newtonsche Analogie}
Diese und "ahnliche Schwierigkeiten haben schon lange vor Einsteins
erster kosmologischer Arbeit dazu gef"uhrt, Ab"anderungen
des Newtonschen Gravitationsgesetzes zu diskutieren. Von diesen
Versuchen werden wir nur einen kurz erw"ahnen, weil er bei der 
Begr"undung des kosmologischen Terms eine gewisse Rolle gespielt 
hat und gerade deshalb Anlass zu Missverst"andnissen geben k"onnte.
Dieser ging aus von dem Astronomen H.~Seeliger, der, aufbauend
auf Vorarbeiten C.~Neumanns, eine zus"atzliche exponentielle 
D"ampfung des "ublichen Potentials in der Form
$\exp(-r\sqrt{\Lambda})/r$ vorschlug~\cite{Seeliger-1909}, 
wobei $\Lambda$ eine fundamentale Konstante von der Dimension
${\hbox{L"ange}}^{-2}$ ist. (Tats"achlich hatte schon Laplace in 
seiner \emph{M\'ecanique C\'eleste} "uber eine universelle 
Absorption der Gravitation in Materie und eine exponentielle 
D"ampfung des Kraftgesetzes spekuliert~\cite{Laplace-1882}.)
Auf der Ebene der Feldgleichungen entspricht dies einer Erg"anzung 
der "ublichen Poissongleichung durch einen zu $\Lambda$ proportionalen 
Term in der Form  
\begin{equation}
\Delta\phi-\Lambda\phi=4\pi G\rho\,,
\label{Seeliger}
\end{equation}
wobei $\phi$ das Gravitationspotential bezeichnet, $\rho$ die 
Massendichte und $G$ die Gravitationskonstante.
Diese Gleichung besitzt nun eine regul"are L"osung f"ur konstante
Materieverteilung $\rho>0$, n"amlich einfach $\phi=-4\pi G\rho/\Lambda$.
Dieser k"onnen dann die Gravitationsfelder lokaler Inhomogenit"aten
einfach "uberlagert werden. Die oben beschriebenen Schwierigkeiten
treten jetzt nicht mehr auf.

Diese Modifikation der Poissongleichung durch den $\Lambda$-Term 
benutzte Einstein in seiner ersten Arbeit \cite{Einstein-1917} als
Analogie f"ur die Einf"uhrung des kosmologischen Terms in 
seinen Feldgleichungen. Diese Argumentation ist dann in den 
klassischen Lehrb"uchern und Monographien "ubernommen worden,
so z.B.~bei von Laue, Weyl, Pauli und Pais' Einstein-Biographie;
stellvertretend sei hier nur auf S.~215 des ber"uhmten 
Enzyklop"adieartikels~\cite{Pauli-2000} von Wolfgang Pauli verwiesen.
Es ist jedoch zu betonen, dass diese formale Analogie physikalisch 
irref"uhrend ist, da der $\Lambda$-Term in der ART keineswegs den 
Effekt hat, die Reichweite der Gravitation zu beschr"anken.
Quantenfeldtheoretisch ausgedr"uckt wird ja durch (\ref{Seeliger}) 
dem Graviton (das die Rolle des "`Photons"' in der Gravitation spielt)
eine Masse von $m=\sqrt{\Lambda}\hbar/c$ zugeordnet, w"ahrend es 
in der Einsteinschen Theorie mit kosmologischem Term masselos bleibt. 
Entsprechend ist auch (\ref{Seeliger}) \emph{nicht} der Newtonsche 
Limes der Einstein-Gleichungen. 
Um diesen zu verstehen, m"ussen wir zun"achst
bemerken, dass in der ART die Quelle des skalaren Gravitationspotentials
nicht allein durch die Massendichte $\rho$ gegeben ist, sondern durch die
Summe $\rho+3p/c^2$, wo $p$ den Materiedruck bezeichnet. Eine 
kosmologische Konstante $\Lambda$ entspricht in der ART aber sowohl
einer Massendichte $\rho_{\Lambda}=\Lambda /8\pi G$, als auch einem 
Druck
% BEGIN FOOTNOTE
$p_{\Lambda}=-\rho_{\Lambda}c^2$.\footnote{Schon in der Speziellen
Relativit"atstheorie gilt ja, dass die tr"age Masse eines K"orpers
von seinem  Spannungszustand abh"angt. Nach dem Prinzip der Gleichheit
von tr"ager und schwerer Masse gilt dies dann auch f"ur letztere.
Dass Druck = - Energiedichte kann man mit der Energieerhaltung
einsehen: Da $\rho_{\Lambda}=$ konst., vergr"o"sert adiabatische
Volumendilatation die innere Energie um $c^2\rho_{\Lambda}\Delta V$,
gleichzeitig wird dabei aber die Arbeit $-p_{\Lambda}\Delta V$ am
System geleistet. Beide Betr"age m"ussen nun gleich sein.}
% END FOOTNOTE
Somit verursacht der $\Lambda$-Term im Newtonschen Limes den
Zusatz $-2\rho_{\Lambda}$ zur Quelle. Damit ist erkl"art, warum
eine \emph{positive} Vakuumenergiedichte trotzdem eine
gravitativ \emph{absto"sende} Wirkung hat.

F"ur nichtrelativistische Materie der Massendichte $\rho_M$
ist der Newtonsche Limes also nicht durch (\ref{Seeliger}) gegeben,
sondern durch
\begin{equation}
\Delta\phi=4\pi G(\rho_M-2\rho_{\Lambda})\,.
\label{Newton-Limes}
\end{equation}
Folgt man Ref. \cite{Heckmann-1968}, so kann man ausgehend von
(\ref{Newton-Limes}) und den Eulergleichungen f"ur die
Materiebewegung die Friedmann-Lema\^{\i}tre-Gleichungen 
(\ref{Friedmann-1},\ref{Friedmann-2},\ref{Friedmann-3}) 
von Anhang~1 innerhalb einer Newtonschen Kosmologie erhalten 
und interpretieren.

\subsection*{2.2 Das Machsche Prinzip}
Ganz wesentlich f"ur Einsteins Motivation, den kosmologischen Term
einzuf"uhren, war seine damalige Vermutung, dass damit das von ihm so
genannte
\emph{Machsche Prinzip} erf"ullt sei, nach dem die Tr"agheitskr"afte
ihre ausschlie"sliche Ursache in der Wechselwirkung mit anderer
Materie haben sollen. Insbesondere d"urften Tr"agheitskr"afte in
einem Universum ohne Materie nicht existieren. Da Tr"agheit und
Gravitation in der ART durch das "Aquivalenzprinzip aber untrennbar
miteinander zusammenh"angen, bedeutet dies, dass die Feldgleichungen
"uberhaupt keine Vakuuml"osungen zulassen d"urfen. Nun ist z.B. die flache
Raum-Zeit der Speziellen Relativit"atstheorie (Minkowskiraum)
zwar Vakuuml"osung der gew"ohnlichen Feldgleichungen, nicht aber
der Gleichungen mit kosmologischem Term. Einstein erkannte aber
auch, dass die Einf"uhrung des kosmologischen Terms allein noch
nicht ausreichte. Ist der Raum n"amlich nicht 
% Begin Footnote
geschlossen\footnote{"`Geschlossen"' ist ein topologischer 
Begriff, der -- etwas ungenau -- besagt, dass der Raum von 
beschr"ankter Ausdehnung  (genau: kompakt) und ohne Rand ist. 
So ist etwa die Erdoberfl"ache (2-Sph"are) oder die Oberfl"ache 
eines Fahrradschlauches (2-Torus) geschlossen. Als "`offen"' 
hingegen bezeichnet man z.B. die unendliche Ebene (oder 
allgemeiner: den $R^n$).},
% End Footnote
so kann durch
Ansetzen willk"urlicher Randbedingungen ein nicht verschwindendes
Gravitationsfeld "`erzeugt"' werden. Das "`Machsche Prinzip"'
erhoffte Einstein nun dadurch zu implementieren, dass er 
nur \emph{r"aumlich geschlossene} Universen als L"osungen seiner 
neuen Feldgleichungen zulie"s. Seine eigene diesen Forderungen 
gen"ugende L"osung~\cite{Einstein-1917} ist das statische 
\emph{Einstein-Universum} mit r"aumlich homogener, druckloser 
Massenverteilung in einem Raum der Topologie der 3-Sph"are $S^3$
(siehe Anhang~1).

Entgegen Einsteins Hoffnung, dass ohne Materie keine solche L"osung
existieren k"onne, gab noch im gleichen Jahr (1917) de~Sitter
eine statische \emph{Vakuum\/}l"osung der Gleichungen mit
kosmologischer Konstante an, in der der Raum ebenfalls geschlossen
$(S^3$) ist. Seiner urspr"unglichen Hoffnung beraubt, meinte Einstein
sp"ater: "`..von dem Machschen Prinzip sollte man eigentlich "uberhaupt
nicht mehr reden"'. Was diejenigen, die sich nicht an diese Anweisung
halten, trotzdem dar"uber reden, kann der Leser in
\cite{Barbour-Pfister-1995} erfahren.

\subsection*{2.3 Von statischen zum dynamischen Universum}
Eine interessante Eigenschaft der de~Sitter-L"osung bestand darin, 
dass sie trotz ihrer (scheinbaren) Statizit"at eine gravitative
Rotverschiebung implizierte, die als "`de~Sitter-Effekt"' bekannt
wurde. Im Gegensatz zur wirklich statischen L"osung Einsteins
stellte sich aber der statische Charakter der de~Sitter-L"osung 
als nur \emph{lokal} g"ultig heraus: global ist das
ganze de~Sitter-Universum ebenfalls einer zeitlichen "Anderung
unterworfen. Diese etwas subtilen mathematischen Sachverhalte 
wurden erst wesentlich durch Arbeiten des Mathematikers Hermann 
Weyl um 1923 gekl"art. Heute wird konsequenterweise die 
de~Sitter-L"osung als Spezialfall einer nichtstatischen  
Kosmologie in der Familie von Friedmann-Lema\^{\i}tre Modellen (vgl.
Anhang~1) verstanden.

Wegbereiter nichtstatischer Kosmologien waren die bahnbrechenden
Arbeiten A.~Friedmanns (1922) und die davon unabh"angig
entstandenen Arbeiten Lema\^{\i}tres (1927), der die kosmologische
Rotverschiebung zum ersten Male als Folge der globalen Expansionsbewegung
interpretierte. Trotzdem dominierten die "`statischen"' L"osungen
Einsteins und de~Sitters die kosmologische Diskussion der 20er-Jahre. 
So hat erstaunlicherweise selbst Hubble noch 1929 die Resultate seiner 
vorhergehenden Beobachtungen der galaktischen Rotverschiebungen im Rahmen
des als statisch angenommenen de~Sitter-Universums interpretiert
("`de~Sitter-Effekt"'),
und Einstein soll im Oktober 1927 anl"asslich des Solvay-Kongresses
in Br"ussel zu Lema\^{\i}tre gesagt haben: "`Vos calculs sont corrects,
mais votre physique est abominable"' (vgl. Sch"uckings Beitrag zu 
\cite{Boerner-et-al-1993} und Kap.~5 in \cite{Gribbin-1999}). 

Schlie"slich setzte sich aber Lema\^{\i}tres erfolgreiche Erkl"arung 
von Hubbles epochaler Entdeckung durch, vor allem unter dem Einfluss
Eddingtons. Auch Einstein stellte sich bald auf die neue Lage ein und
verwarf den kosmologischen Term und damit nat"urlich auch seine alte
statische L"osung; nicht nur weil letztere keine Rechenschaft
der kosmologischen Rotverschiebung geben konnte, sondern vor allem
auch deshalb, weil sie sich als gegen"uber St"orungen instabil 
erwies~\cite{Einstein-1931} (siehe Anhang~1).

\subsection*{2.4 Das weitere Schicksal Lambdas}
Einer weit verbreiteten "Uberlieferung zufolge soll Einstein 
sp"ater seine Einf"uhrung des kosmologischen Terms "`als gr"o"ste 
Dummheit meines Lebens"' bezeichnet haben; ein Beleg f"ur dieses 
Zitat ist uns allerdings unbekannt. In \cite{Einstein-1956} schrieb 
er dazu: "`W"urde die Hubble--Expansion bei Aufstellung der
Allgemeinen Relativit"atstheorie bereits entdeckt gewesen sein,
so w"are es nie zur Einf"uhrung des kosmologischen Terms gekommen.
Es erscheint nun a~posteriori um so ungerechtfertigter, einen solchen 
Term in die Feldgleichungen einzuf"uhren, als dessen Einf"uhrung 
seine einzige urspr"ungliche Existenzberechtigung -- zu einer 
nat"urlichen L"osung des kosmologischen Problems zu f"uhren -- 
einb"u"st."' Viele der einflussreichen Kommentatoren, so z.B.
Pauli und Jordan (siehe stellvertretend \cite{Pauli-2000},
S.~265 Anm.~18) haben sich diesem Urteil angeschlossen; doch
gab es auch Ausnahmen (siehe z.B. \cite{Heckmann-1968}, Anm. zu
S.~35).

Zeitweise geriet das Standardmodell von Friedmann und Lema\^{\i}tre
ohne kosmologischen Term in Misskredit, da das Alter des Universums
aufgrund von Hubbles Messungen im Vergleich zum Alter der Sterne als
zu kurz herauskam. Deshalb wurde der $\Lambda$--Term wieder
eingef"uhrt, und ein fr"uheres Modell von Lema\^{\i}tre mit
verz"ogerter Expansion erfuhr eine Wiederbelebung. Das "anderte
sich aber, als neue astronomische Beobachtungen, vor allem durch W.~Baade
am Mt.~Palomar-Observatorium, zu einer erheblichen Revision des
Hubble--Parameters f"uhrten. Eine ausf"uhrliche Schilderung 
dieser spannenden Entwicklung findet man in \cite{Gribbin-1999}. 

In j"ungerer Zeit haben nun relativ gro"se Werte des Hubble-Parameters,
die von gewissen Gruppen vertreten wurden, eine erneute "`Alterskrise"'
verursacht, zu deren Behebung als Medizin einmal mehr den kosmologischen
Term herbeizitiert wurde. Auch andere Gr"unde f"ur den $\Lambda$-Term 
wurden ins Feld gef"uhrt, welche sich aber bald -- Dank erweiterter 
Beobachtungen -- als voreilig erwiesen. Die neuesten experimentellen 
Daten scheinen nun viel "uberzeugender als bisher erneut einen positiven
Wert der kosmologischen Konstante zu fordern. Es bleibt gespannt abzuwarten, 
ob es diesmal dabei bleiben wird.

\section*{3 Das Problem der kosmologischen Konstante}
Eine charakteristische Massendichte der Kosmologie ist die 
sogenannte \emph{kritische Dichte}
\begin{equation}
\rho_{\rm krit}:=
\frac{3 H_0^2}{8\pi G}\approx 10^{-29}\frac{g}{cm^3}\,,
\label{def-rho-krit}
\end{equation}
die im Falle einer verschwindenden kosmologischen Konstante
einer flachen Geometrie zugeh"orig ist, w"ahrend gr"o"sere
bzw. kleinere Massendichten dann zu positiven bzw. negativen 
Kr"ummungen f"uhren. Auch im allgemeinen Fall ($\Lambda\not =0$),
in dem der Zusammenhang von Massendichte und Kr"ummung noch 
$\Lambda$ involviert (vgl. (\ref{triangle}) in Anhang~1), beh"alt 
man $\rho_{\rm krit}$ als Bezugsgr"o"se bei.

In (\ref{def-rho-krit}) ist $G$ die Newtonsche Gravitationskonstante 
und $H_0$  der \emph{Hubble-Parameter} zum gegenw"artigen Zeitpunkt 
(deshalb Index $0$). F"ur nicht allzu ferne Galaxien ist deren
Geschwindigkeit proportional zur Entfernung. 
Die Proportionalit"atskonstante in diesem \emph{Hubbleschen Gesetz}
ist gerade der Hubble-Parameter, der die Dimension einer inversen 
Zeit hat. Der Wert von $H_0$ bestimmt im Wesentlichen auch alle anderen 
kosmologischen Skalen und wird gerne in Form der sogenannten 
"`reduzierten Hubble-Konstante"' 
$h_0:=H_0/100\cdot Km\cdot s^{-1}\cdot Mpc^{-1}$ angegeben, deren 
Wert gegenw"artig auf $h_0\approx 0,65\pm 0,1$ eingegrenzt ist;
dabei ist $1Mpc=3,086\cdot 10^{19}Km$.

%%%%%%%%%%%%%%% ABB 1 %%%%%%%%%%%%%%%%%%%%%%%%%%%%%%%%%%%%%%%%

\begin{figure}
\includegraphics[width=12cm]{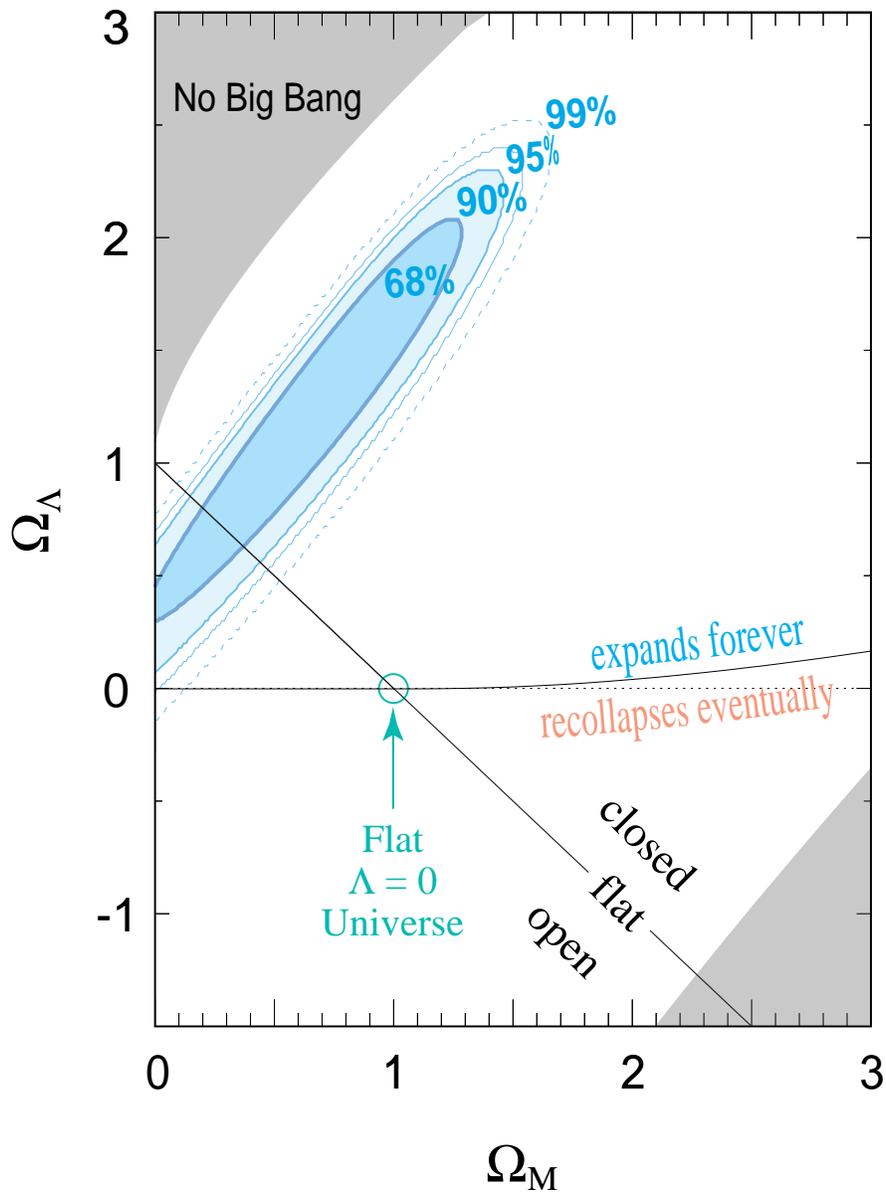}
\caption{\it
Konfidenzgebiete in der $(\Omega_M,\Omega_{\Lambda})$ -- Ebene,
wie sie sich aus den Supernovae-Daten ergeben. Das schattierte
Gebiet links oben entspricht zur"uckprallenden Modellen ohne
Urknall. Das schattierte Gebiet rechts unten ist ausgeschlossen,
da das Alter des Universums in diesem Bereich zu kurz ist.
Oberhalb der (nicht eingezeichneten) Geraden
$\Omega_{\Lambda}=\frac{1}{2}\Omega_M$
verl"auft die Expansion beschleunigt, unterhalb verz"ogert.
Das Diagramm ist der Homepage des SCP \cite{Home-SCP}
entnommen.}
\label{fig:Abb1}
\end{figure}           

%%%%%%%%%%%%%%%%%%%%%%%%%%%%%%%%%%%%%%%%%%%%%%%%%%%%%%%%%%%%%%

Verschiedene Beobachtungen zeigen nun, dass die der kosmologischen 
Konstante entsprechende gravitierende Energiedichte 
$\rho_{\Lambda}c^2$ nicht wesentlich gr"o"ser als 
$\rho_{\rm krit}c^2$ sein kann. Zum Einen w"urde n"amlich ein zu 
gro"ses $\rho_{\Lambda}$ (vgl. 
%
% Abb. 1. zuerst genannt 
%
die linke obere Ecke von Abbildung~\ref{fig:Abb1}) den Urknall 
%% BEGIN FOOTNOTE
verhindern\footnote{ 
Genauer gilt im uns interessierenden Wertebereich 
$0,1\leq\Omega_M\leq 1$, $\vert\Omega_{\Lambda}\vert\leq 1$, 
dass das Weltalter divergiert, falls $\Omega_{\Lambda}$
eine kritische obere Grenze von etwa $1+\frac{7}{3}\Omega_M$
"uberschreitet.  Das ergibt sich sofort aus Gl.~(5.3) von
\cite{Peacock-1999}.}
%% END FOOTNOTE
und damit unsere Erkl"arung der H"aufigkeit leichter Elemente 
zunichte machen. Zum Anderen setzen Altersbestimmungen an 
Kugelsternhaufen eine obere Grenze an das Weltalter und damit an 
$\rho_{\Lambda}$, da ja ein positives $\rho_{\Lambda}$ den Effekt 
hat, die Expansionsgeschwindigkeit in der Vergangenheit zu 
verlangsamen und damit das Weltalter zu erh"ohen.
In den der Teilchenphysik angemessenen Einheiten ist diese 
Schranke infinitesimal:
\begin{equation}
\rho_{\Lambda}c^2\leq 10^{-46} (GeV)^4/(\hbar c)^3\,.
\label{upper-bound}
\end{equation}
Wie bereits erw"ahnt ist es n"amlich mehr als r"atselhaft, weshalb
aufgrund von
Quantenfluktuationen das Standardmodell der Elementarteilchenphysik 
nicht eine gravitativ wirksame Vakuumenergiedichte nach sich zieht, 
welche die zuletzt angegebene Schranke gewaltig "uberschreitet. 
Prinzipielles dazu wird in siehe Anhang~2 ausgef"uhrt. 
Als Ma"sstab kann z.B. die 
\emph{Kondensationsenergiedichte} beim Phasen"ubergang der 
Quantenchromodynamik dienen, die etwa $10^{-1} (GeV)^4/(\hbar c)^3$
betr"agt und somit die angegebene Schranke um mehr als 
40 Gr"o"senordnungen "ubersteigt. F"ur denkbare Kompensationen 
verschiedener Beitr"age gibt es aber keinen Grund, da alle Symmetrien 
-- au"ser der elektromagnetischen Eichgruppe -- spontan gebrochen sind. 
Zwar ist es logisch m"oglich, neben dem durch die
Vakuumfluktuationen induzierten $\Lambda_{\rm vac}$ ein in den 
Feldgleichungen a priori vorhandenes $\Lambda_0$ anzunehmen, so dass
dann deren Summe -- das effektive $\Lambda$ -- die experimentelle 
Schranke nicht "ubersteigt. Dieses $\lambda_0$ entspr"ache dann
keiner neuen Energieform, sondern w"are als gleichberechtigter 
Bestandteil der Vakuum-Einsteingleichungen anzusehen.  
Doch w"urde eine solch extreme a priori 
Anpassung der Betr"age ("`fine-tuning"') bis auf 40 Stellen keinen 
physikalischen Erkl"arungswert besitzen. Wir stehen hier vor einem 
tiefen R"atsel, zu dessen L"osung uns vorl"aufig die Grundlagen 
fehlen. Mehr dazu findet man im "Ubersichtsartikel von 
S.~Weinberg~\cite{Weinberg-1989}. 

Die Frage nach einer gravitativ wirksamen Vakuumenergiedichte wurde 
von Pauli schon in seinen jungen Jahren anhand eines einfachen Falls 
gestellt: Er fragte sich, ob vielleicht die Nullpunkts\-energie des  
elektromagnetischen Feldes ein Gravitationsfeld erzeugen k"onnte.
Wegen der unendlich vielen Freiheitsgrade ist diese zwar divergent, 
aber jede Physikergeneration hat ihre Abschneideskala, und das war 
damals der klassische Elektronenradius ($R_e=2,8\cdot 10^{-15}m$),
entsprechend einer maximalen Modenenergie ($2\pi\hbar c/R_e$) von 
etwa $0,5\,GeV$. 
Die so gewonnene Energiedichte setzte Pauli versuchsweise in die
Grundgleichung f"ur das statische Einstein-Universum ein und 
stellte am"usiert fest, dass "`das Universum nicht einmal bis zum 
Mond reicht"' (berichtet in~\cite{Enz-Thellung-1960}); tats"achlich 
ergibt die Rechnung einen Radius von nur $32\,Km$.
Heute w"urden wir die effektive Abschneideskala f"ur die 
maximale Modenenergie sogar noch viel h"oher ansetzen, n"amlich 
bei der sogenannten Fermi-Skala von etwa $100~GeV$, was den 
prognostizierten Radius des Einstein-Universums weiter auf 
$13\,cm$ schrumpfen lie"se, etwa 27 Gr"o"senordnungen unter
dem heutigen Hubble-Radius. (Dieser entspricht etwa der Distanz 
\emph{Weltalter $\times$ Lichtgeschwindigkeit}, wobei das Weltalter 
ungef"ahr bei $1,5\cdot 10^{10}$ Jahren liegt.) 
Damit d"urfte es sich um die schlechteste Prognose in der 
Geschichte der Physik handeln. 

\section*{4 Hubble-Diagramm der Supernovae vom Typ Ia bei 
          hohen Rotverschiebungen}
Naiv versteht man zun"achst unter einem Hubble-Diagramm die Kurve,
in der die Fluchtgeschwindigkeit als Funktion des Abstandes aufgetragen 
ist. Diese ist, wie schon gesagt, in sehr guter N"aherung eine Gerade 
durch den Ursprung deren Steigung durch den Hubble-Parameter $H_0$ 
gegeben ist. Verl"auft die kosmische Expansion beschleunigt oder 
verz"ogert, so werden aber f"ur gro"se Abst"ande Abweichungen 
von der Geraden dadurch eintreten, dass die hier und heute bestimmten 
Geschwindigkeiten dem betreffenden Objekt um die Lichtlaufzeit fr"uher 
zukommen und deshalb entsprechend kleiner bzw. gr"o"ser sind.

Es ist jedoch zu bedenken, dass f"ur die gro"sen Skalen, mit denen
es die Kosmologie zu tun hat, weder Abst"ande noch Geschwindigkeiten 
direkt messbare Gr"o"sen sind. Eine als Abstandsma"s verwendete
Gr"o"se ist die sogenannte \emph{Helligkeitsdistanz} $D_L$, die wir
gleich vorstellen werden. Anstatt der Geschwindigkeit benutzt man die
Rotverschiebung $z$, die durch $\lambda_e=(1+z)\lambda_a$ definiert 
ist, wobei $\lambda_e$ die empfangene, $\lambda_a$ die ausgestrahlte
Wellenl"ange ist. 
Es ist an dieser Stelle zu betonen, dass auch aus prinzipiellen 
Erw"agungen der Begriff der Flucht-\emph{Geschwindigkeit} hier sehr
problematisch ist, da wir es mit weit entfernten Objekten in einer
\emph{zeitlich ver"anderlichen Geometrie} zu tun haben. Nicht die
Objekte bewegen sich \emph{im Raum} (lokal sind sie im Mittel in Ruhe), 
sondern der Raum zwischen ihnen dehnt sich aus. Es ist daher i.A. auch 
sinnlos, die Rotverschiebung nach der Dopplerschen Formel in eine
Geschwindigkeit zur"uckzurechnen, was zu allerlei Fehlinterpretationen 
Anlass geben kann.  

Tr"agt man nun diese beiden Gr"o"sen gegeneinander, mitunter mit
vertauschten Achsen, also etwa die Helligkeitsdistanz (bzw. eine
logarithmische Funktion von ihr) als Funktion der Rotverschiebung,
so erh"alt man das, was man heute allgemein ein Hubble-Diagramm nennt.

Bevor die neuen aufregenden Daten f"ur die Klasse der Typ~Ia 
Supernovae vorgestellt und diskutiert werden, m"ussen wir kurz 
einige Fakten wiederholen, um zu verstehen welche Gr"o"sen
eigentlich gemessen wurden.

\subsection*{4.1 Theoretischer Hintergrund}
In der Kosmologie werden unterschiedliche Distanzma"se verwendet,
welche aber "uber einfache Rotverschiebungsfaktoren verkn"upft sind. 
F"ur das Folgende ist die \emph{Helligkeitsdistanz} $D_L$ ma"sgebend,
die durch $D_L=\sqrt{L/4\pi F}$ definiert ist, wo $L$ die 
intrinsische  Luminosit"at der Quelle und $F$ der beobachtete 
Energiefluss ist. Wie bei allen kosmologischen Distanzen setzt man 
$D_L$ zweckm"a"sigerweise proportional zu $c/H_0=3000h_0^{-1} Mpc$
und erh"alt
\begin{equation}
D_L(z)=\frac{c}{H_0}d_L(z;\Omega_M,\Omega_{\Lambda})
\label{def-d_L}
\end{equation}
mit der dimensionslosen Gr"o"se $d_L$, die in allen 
Friedmann-Lema\^{\i}tre-Modellen eine bekannte Funktion der 
Rotverschiebung $z$ und der wichtigen kosmologischen Parameter
\begin{equation}
\Omega_M:=\frac{\rho_M}{\rho_{\rm krit}},\quad
\Omega_{\Lambda}:=\frac{\rho_{\Lambda}}{\rho_{\rm krit}}
\label{def-omega}
\end{equation}
ist. Dabei repr"asentiert $\Omega_M$ die gesamte Materie, 
einschlie"slich der dunklen Materie samt ihrem nichtbaryonischen 
Anteil. Die Gr"o"se $\Omega_K=1-\Omega_{M}-\Omega_{\Lambda}$ 
ist ein Ma"s f"ur die (r"aumlich konstante) Kr"ummung des Raumes:
Bei positiver bzw. negativer Kr"ummung ist $\Omega_K$ negativ bzw. 
positiv. F"ur einen flachen Raum ist $\Omega_K=0$, d.h. 
$\Omega_M+\Omega_{\Lambda}=1$ (siehe Anhang~1). 

Astronomen benutzen als logarithmische Ma"se f"ur $L$ und $F$ 
sogenannte \emph{absolute} bzw. \emph{scheinbare Magnituden}
$M$ und $m$. Die Konventionen sind dabei so getroffen, dass
der \emph{Distanzmodul} $m-M$ mit $D_L$ so zusammenh"angt:
\begin{equation}
m-M= 5\,\log\left(\frac{D_L}{1 Mpc}\right)+25\,.
\label{distanzmodul}
\end{equation}
Wird hier die obige Darstellung von $D_L$ eingesetzt, so ergibt sich  
die folgende Beziehung zwischen scheinbarer Magnitude $m$ und 
Rotverschiebung~$z$:
\begin{equation}
m=\mu+5\,\log d_L(z;\Omega_M,\Omega_{\Lambda})\,,
\label{m-z-Beziehung}
\end{equation}
mit $\mu=M-5\,\log(H_0\cdot 1Mpc/c)+25$. F"ur Standardkerzen 
(einheitliches $M$) ist dies nat"urlich eine Konstante, welche als  
Anpassungsparameter behandelt wird. Durch Vergleich dieser
theoretischen Erwartung mit Beobachtungsdaten werden sich interessante 
Einschr"ankungen an die kosmologischen Parameter $\Omega_M$ und 
$\Omega_{\Lambda}$ ergeben.

In diesem Zusammenhang ist die folgende Bemerkung wichtig: Halten wir 
$z$ im relevanten Intervall zwischen $0,4$ und $0,8$ der derzeitigen 
Beobachtungen fest, so definieren die Gleichungen 
$d_L(z;\Omega_M,\Omega_{\Lambda})=\hbox{konst.}$ Entartungskurven
in der $\Omega$-Ebene. Da die Kr"ummungen dieser Kurven sich 
als gering erweisen, k"onnen wir ihnen eine ungef"ahre Steigung 
zuordnen. F"ur $z=0,4$ ist diese nahe bei 1 und 
nimmt auf $1,5$ bis $2$ bei $z=0,8$ zu, f"ur den ganzen 
interessierenden Wertebereich von $\Omega_M$ und $\Omega_{\Lambda}$. 
Deshalb k"onnen auch sehr genaue Daten nur ein schmales 
Band in der $\Omega$-Ebene aussondern, was wiederum die langgestreckte 
%
% Abbildung 1 zum 2. mal
%
Form der "`Likelihood-Gebiete"' erkl"art, die in 
Abbildung~\ref{fig:Abb1} gezeigt
werden. Er wird also darauf ankommen, mit unabh"angigen Methoden 
mehrere, sich m"oglichst transversal schneidende Gebiete auszusondern.

\subsection*{4.2 Supernovae vom Typ Ia als Standardkerzen}
Schon in den drei"siger Jahren wurde erkannt, dass Supernovae vom 
Typ~Ia ausgezeichnete Standardkerzen sind, welche zudem bis zu 
kosmischen Distanzen von $\approx 500\,Mpc$ sichtbar 
sind~\cite{Baade-1938}. F"ur n"ahere Distanzen eignen sie sich 
besonders zur Bestimmung des Hubble-Parameters. Daf"ur ist freilich 
eine \emph{Kalibrierung} der absoluten Magnitude mit verschiedenen     
Distanzbestimmungen n"otig. Dank des Weltraumteleskops \emph{Hubble} 
(HST) wurde dies mit Hilfe von sogenannten Cepheiden
m"oglich~\cite{Parodi-et-al-2000}. Cepheiden sind pulsierende und damit 
periodisch ihre Helligkeit ver"andernde Sterne, deren absolute
Helligkeit in einem ann"ahernd festen Verh"altnis zu ihrer
Pulsationsfrequenz steht. Beobachtungen dieser Frequenz und der 
scheinbaren Helligkeit lassen daher R"uckschl"usse auf die Entfernung 
zu.

Wie eingangs erw"ahnt, ergeben sich bei beschleunigter bzw. 
verz"ogerter Expansion Abweichungen vom linearen Hubble-Gesetz
bei gro"sen Abst"anden. Diese werden mit dem sogenannten 
\emph{Bremsparameter} charakterisiert (siehe Anhang~1).
Daher haben bereits 1979 Tammann~\cite{Tammann-1979} und  
Colgate~\cite{Colgate-1979} unabh"angig voneinander vorgeschlagen, 
dass die genannte Klasse von Supernovae bei h"oheren Rotverschiebungen 
zur Bestimmung dieses Bremsparameters benutzt werden k"onnte. 
Inzwischen ist die Verwirklichung dieses Programms Dank neuer 
Technologien erm"oglicht worden. Wesentlich f"ur die Entdeckung der  
Supernovae bei hohen Rotverschiebungen sind gro"sfl"achige Detektoren
an Gro"steleskopen, die es erm"oglichen, digitale Aufnahmen relativ 
gro"ser Himmelsareale zu erhalten. F"ur die anschlie"senden 
photometrischen und spektroskopischen Untersuchungen ist ferner 
der Einsatz der besten Gro"steleskope, wie dem \emph{HST} und 
\emph{Keck}, unerl"a"slich.

Zwei Forschergruppen haben diese Untersuchungen in den letzten Jahren 
vorangetrieben, n"amlich das "`Supernovae Cosmology Project"' (SCP,
Homepage \cite{Home-SCP})
und das "`High-Z Supernova Search Team"' (HZT). Beide Gruppen haben 
je "uber 70 Supernovae vom Typ~Ia entdeckt und publizierten in j"ungster 
Zeit fast identische Ergebnisse; siehe \cite{Perlmutter-et-al-1999}
bzw. \cite{Riess-et-al-1999}. In Anbetracht deren Bedeutung ist diese 
"Ubereinstimmung besonders hervorzuheben.

Bevor wir auf die Resultate eingehen, sind noch einige, zum Teil 
kritische Bemerkungen "uber die Natur und die physikalischen 
Eigenschaften einer Supernova vom Typ~Ia n"otig. Ihr unmittelbarer 
Vorg"anger ist wahrscheinlich ein 
Weisser Zwerg, welcher haupts"achlich aus Kohlenstoff und Sauerstoff 
besteht und Teil eines engen Doppelsternsystems ist. Im Standardszenario 
str"omt vom Begleiter Materie auf den Weissen Zwerg, so dass unter 
Umst"anden dessen Masse stetig zunimmt, bis die kritische 
Chandrasekhar-Masse erreicht wird und der Weisse Zwerg instabil wird.  
Dies f"uhrt dann entweder zu einem Kollaps auf einen Neutronenstern 
oder zur Z"undung des entarteten Kohlenstoffs tief im Inneren.
Im zweiten Fall entsteht eine nach au"sen propagierende subsonische 
nukleare Brennfront (eine Deflagration), wobei der Weisse Zwerg 
vollst"andig zerrissen wird. In wenigen Sekunden wird das Sternmaterial 
weitgehend in Nickel und andere Elemente zwischen Silizium und Eisen 
umgewandelt. Das ausgesto"sene Nickel zerf"allt in Kobalt und danach 
in Eisen. Es muss aber betont werden, dass die Physik dieser 
thermonuklearen Explosion von entarteter Materie sehr komplex
und noch nicht hinreichend verstanden ist. Deshalb sind auch die 
durchgef"uhrten numerischen Simulationen im Detail noch nicht als 
zuverl"assig anzusehen.

Supernovae des Typs Ia sind keine \emph{perfekten}
Standardkerzen. Ihre Maximalleuchtkr"afte zeigen je nach Auswahl
eine Dispersion von $0,3 - 0,5\, mag$. Dabei bezeichnet $mag$ 
das in der messenden Astronomie benutzte Helligkeitsma"s, welches mit dem
Intensit"atsverh"altnis $I_1/I_2$ so zusammenh"angt:
$\log(I_1/I_2)=-0,4(m_1-m_2)$. Die Astronomen haben jedoch 
gelernt, die Unterschiede herauszukorrigieren und somit 
die intrinsische Dispersion unter $0,17\, mag$ zu dr"ucken. 
Dabei benutzen sie eine intrinsische Korrelation zwischen 
maximaler Helligkeit und Breite der Lichtkurve, die es 
gestattet, vermittels einfacher Streckungen der Zeitachse 
alle Lichtkurven mit erstaunlicher Genauigkeit zur Deckung 
zu bringen. Nat"urlich m"ussen auch andere Korrekturen,
wie z.B. die galaktische Extinktion in Rechnung gestellt 
werden. Am Ende wird f"ur jede Supernova eine effektive 
scheinbare Magnitude $m_B^{\rm eff}$ im blauen Filter (bez"uglich
des Ruhesystems) bestimmt. Die Abh"angigkeit dieser Gr"o"se von der 
Rotverschiebung wird schlie"slich mit der theoretischen Erwartung
(\ref{m-z-Beziehung}) verglichen.

\subsection*{4.3 Ergebnisse}
%
% ABB 2
%
\begin{figure}
\includegraphics[width=12cm]{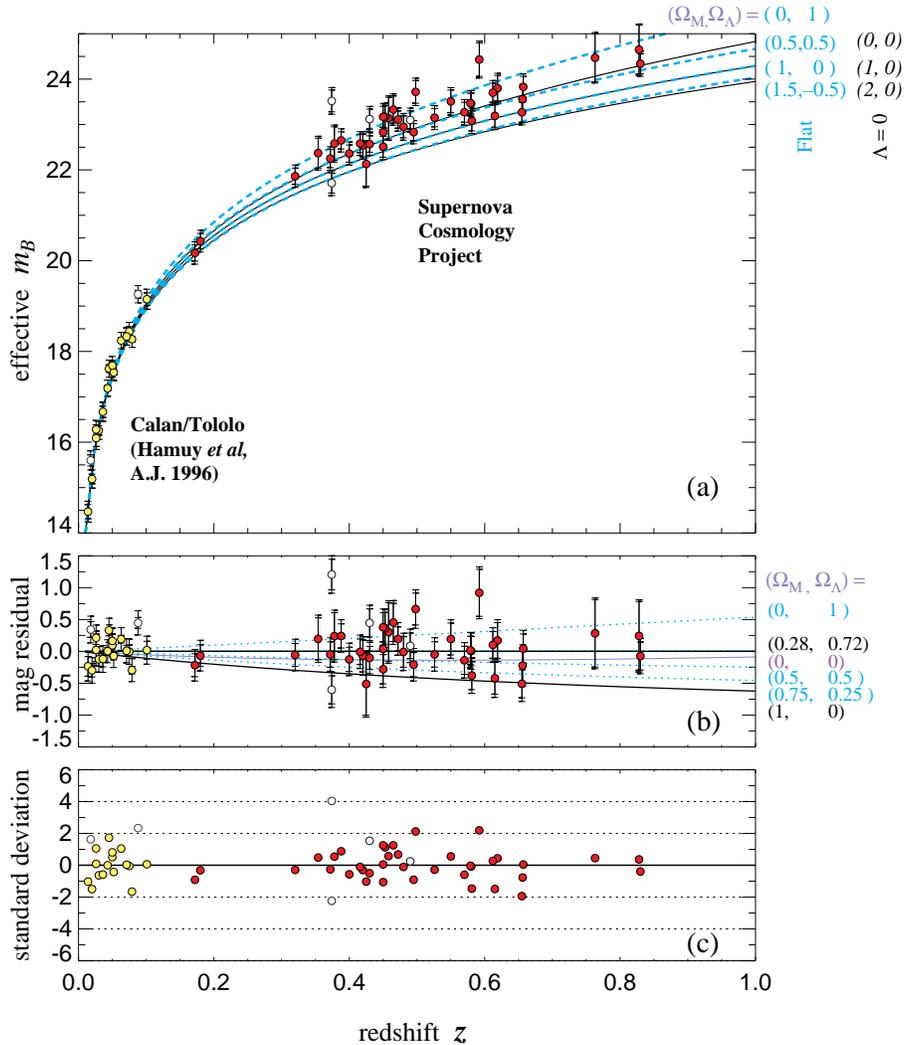}
\caption{\it
Oberer Kasten:
Magnituden-Rotverschiebungs-Beziehung (Hubble-Diagramm) f"ur
42 Typ-Ia-Supernovae bei hohen Rotverschiebungen des SCP,
zusammen mit 18~Supernovae bei kleinen Rotverschiebungen der
Cal\'an/Tololo - Durchmusterung. Die durchgezogenen Kuven
entsprechen den theoretischen Erwartungen f"ur $\Lambda=0$,
die gestrichelten denjenigen f"ur verschiedene flache
Modelle.
Mittlerer Kasten:
Abweichungen der Magnituden zum besten Fit
$(\Omega_M,\Omega_{\Lambda})=(0.28,0.72)$ unter den flachen
Modellen.
Unterer Kasten:
Die Abweichungen scheinen ohne erkennbare Tendenz
zuf"allig zu streuen. Das Diagramm ist der Homepage
des SCP~\cite{Home-SCP} entnommen.}
\label{fig:Abb2}
\end{figure}         
%%%%%%%%%%%%%%%%%%%%%%%%%%%%%%%%%%%%%%%%%%%%%%%%%%%%%%%%%%
%
% Abb. 2 zuerst genannt
%
Abbildung~\ref{fig:Abb2} zeigt das Hubble-Diagramm f"ur $m_B^{\rm eff}$
der 42 Typ-Ia-Supernovae bei hoher Rotverschiebung, das von der
SCP-Gruppe publiziert wurde \cite{Perlmutter-et-al-1999}, 
zusammen mit 18 Cal\'an/Tololo Supernovae bei kleinen
Rotverschiebungen. (F"ur die Ergebnisse des HZT siehe 
\cite{Riess-et-al-1999}.)

% Abb. 1 zum 3. mal genannt
%
Das haupts"achliche Resultat der Analyse ist in 
Abbildung~\ref{fig:Abb1}
dargestellt. Die Lage der Konfidenzgebiete in der 
$(\Omega_M,\Omega_{\Lambda})$ -- Ebene zeigt insbesondere, dass 
$\Omega_{\Lambda}$ auf dem $95\%$-Konfidenzniveau  
von Null verschieden und positiv 
%% BEGIN FOOTNOTE
ist.\footnote{Die vollst"andige
Analyse in \cite{Perlmutter-et-al-1999} ergibt sogar ein
positives $\Omega_{\Lambda}$ auf dem $99\%$-Konfidenzniveau. 
Darauf brauchen wir aber nicht n"aher einzugehen,
da die neuen, noch zu besprechenden Mikrowellendaten kleine
$\Omega_{\Lambda}$ ebenfalls ausschlie"sen; siehe 
Abbildung~\ref{fig:Abb3}.}     
%% END FOOTNOTE
Eine approximative Anpassung ist
\begin{equation}
0,8\Omega_M-0,6\Omega_{\Lambda}\approx -0,2\pm 0,1\,.
\label{Anpassung}
\end{equation}

Da die l"anglichen "`Likelihood-Gebiete"' ann"ahernd senkrecht 
auf den Geraden konstanter Raumkr"ummung 
($\Omega_M+\Omega_{\Lambda}=\hbox{konst.}$) stehen, sagen diese 
Messungen fast nichts "uber die r"aumlichen Kr"ummungsverh"altnisse.
Diese werden erst durch die unten zu besprechenden Messungen der
Anisotropie des Mikrowellenhintergrundes wesentlich eingeschr"ankt, deren 
"`Likelihood-Gebiete"' stark entlang der Geraden verschwindender 
Raumkr"ummung konzentriert sind. Setzt man also die Raumkr"ummung mit
null an, so muss nach Gleichung (\ref{triangle}) aus Anhang~1 {}
$\Omega_M+\Omega_{\Lambda}=1$ gelten. Damit ergibt sich dann der 
Materieanteil der Massendichte, einschlie"slich dunkler 
(auch nichtbaryonischer) Materie, zu 
\goodbreak
\begin{eqnarray}
\Omega_M^{\rm flach} = 0,28
&&\hspace{-0.6cm}
{}^{+0,09}_{-0,08}\,\hbox{($1\sigma$ statistisch)}
\qquad\nonumber\\   
&&\hspace{-0.6cm}
{}^{+0,05}_{-0,04}\,\hbox{(identifizierte syst. Fehler)}.   
\qquad
\label{Omega-M}
\end{eqnarray}

\subsection*{4.4 Systematische Fehler}
Hinsichtlich der Frage der Vorl"aufigkeit dieser Ergebnisse
d"urfen Hinweise auf m"ogliche Fehlerquellen selbstverst"andlich
nicht unterdr"uckt werden. Die Ergebnisse beider Forschungsgruppen
sind in erster Linie durch \emph{systematische Fehler} begrenzt,
die eingehend untersucht werden.
Nach Ber"ucksichtigung
aller bisher quantifizierten Fehler scheint es unm"oglich zu 
sein, die Daten mit $\Omega_{\Lambda}=0$ zu beschreiben. Trotzdem 
werden noch viele Untersuchungen n"otig sein, um sicherzugehen, 
%
% Abb 1 zum 4. mal genannt
%
dass die in Abbildung~\ref{fig:Abb1} gezeigten Resultate nicht doch durch 
systematische Effekte verf"alscht werden. Man wird weiter nach
unidentifizierten Effekten suchen, aufgrund deren die Supernovae 
bei hohen Rotverschiebungen systematisch um etwa 15\% schw"acher 
erscheinen w"urden.

Neben Extinktion durch die Muttergalaxie und im intergalaktischen 
Raum beunruhigen am meisten evolutive Effekte, die sich 
daraus ergeben k"onnen, dass die von uns in sehr gro"sen
Entfernungen wahrgenommenen Supernovae-Ereignisse zu einer sehr 
viel fr"uheren Evolutionsphase des Universums stattfanden, in der
die verschiedenartigen (z.B. metall"armere) chemischen H"aufigkeiten 
auch zu Variationen der Explosionsvorg"ange f"uhren k"onnen. 
Solche Effekte sind zweifellos 
vorhanden und werden auch theoretisch studiert. Durch das 
einfache Streckungsverfahren scheinen diese jedoch weitgehend 
korrigiert zu werden, denn es verbleibt danach kein evolutiver 
Trend in den Lichtkurven und auch die Spektren "andern sich nicht. 
All dies muss aber noch weiter untersucht werden. Besonders wichtig
ist die Ausdehnung der Beobachtungen zu h"oheren Rotverschiebungen.
F"ur eine  detaillierte Diskussion sei der Leser auf die 
Originalarbeiten \cite{Perlmutter-et-al-1999} und 
\cite{Riess-et-al-1999}, sowie auf den k"urzlich erschienenen
"Ubersichtsartikel \cite{Riess-2000} verwiesen.

\section*{5 Anisotropien der Mikrowellenstrahlung}
%
% %%%%%% ABB 3 %%%%%%%%%%%%%%%%%%%%%%%%%%%%%%%%
%
\begin{figure}
\hspace{-1.5cm}
\includegraphics[width=10cm, angle=90]{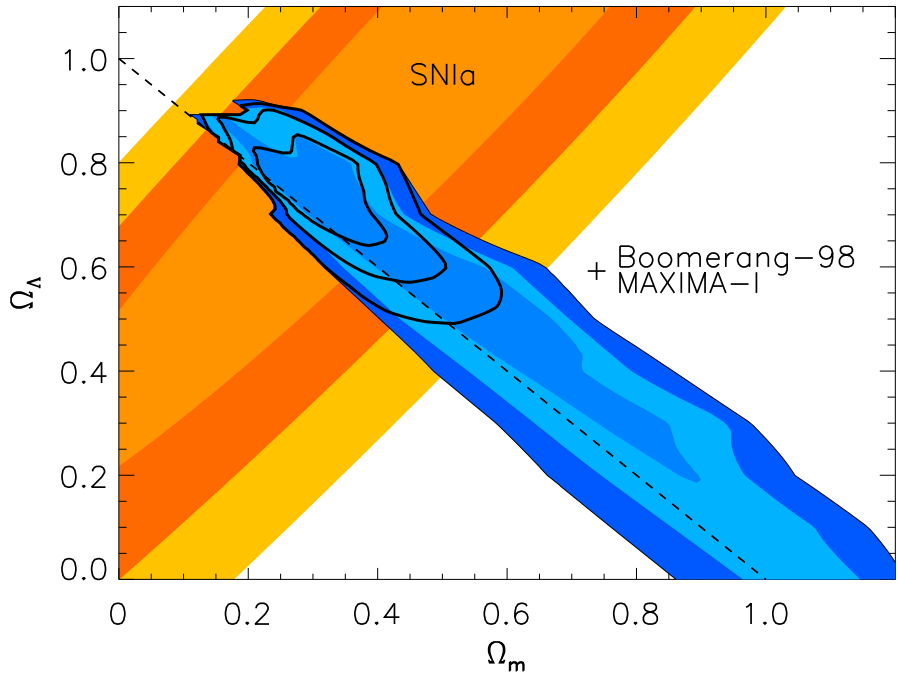} 
\caption{\it
Konfidenzgebiete ($68\%,\,95\%,\,99\%)$ in der
$(\Omega_M,\Omega_{\Lambda})$ -- Ebene
der kombinierten BOOMERANG-98 und MAXIMA-1 Daten (blau) 
mit den Supernova Daten (orange). Es ist deutlich zu sehen, wie 
diese zueinander transversale Gebiete mit einem relativ kleinen
"Uberlappungsgebiet aussondern. Die schwarzen Konturen im
"Uberlappungsgebiet beranden  die entsprechenden Konfidenzgebiete 
der gemeinsamen Daten. Flache Modelle liegen 
auf der schwarz-gestrichelten Diagonalen. 
Die in \cite{Bahcall-et-al-1999} berichteten Daten f"ur
Galaxienhaufen, die hier nicht eingezeichnet sind, sondern einen
weiteren, vertikalen Streifen $\Omega_M=0,3\pm0,1\, (1\sigma)$
aus, der dieses "Uberlappungsgebiet  voll trifft. Das Diagramm ist
\cite{Jaffe-et-al-2000} entnommen.}
\label{fig:Abb3}
\end{figure}       

%%%%%%%%%%%%%%%%%%%%%%%%%%%%%%%%%%%%%%%%%%%%%%%%%%%%%%%%%%%%%%
Obwohl schon die Supernovae-Daten bereits ziemlich deutlich f"ur 
eine positive Vakuumenergiedichte sprechen, wird die Evidenz daf"ur 
wesentlich st"arker in Verbindung mit den neuesten Daten 
"uber die \emph{Anisotropie der kosmischen 
Mikrowellenhintergrundstrahlung}. Das von diesen Daten ausgesonderte 
l"angliche Gebiet in der $\Omega_M,\Omega_{\Lambda})$ Ebene steht 
n"amlich nahezu senkrecht auf demjenigen der Supernovae-Daten und 
ist, wie schon gesagt, um die Gerade zu verschwindender
r"aumlicher Kr"ummung konzentriert. Es ergibt sich damit eine 
%
% Abb 3 zum 1. mal genannt
%
relativ kleines "Uberlappungsgebiet, das in Abbildung~\ref{fig:Abb3}
gezeigt ist. 
 
Die "uberaus wichtige und fruchtbare Erforschung der Anisotropien 
in der kosmischen Mikrowellenhintergrundstrahlung (CMB) hat 
gegenw"artig ein kritisches Stadium erreicht und wird in den kommenden 
Jahren sicherlich eines der dominierenden Themen in der Kosmologie 
bleiben. Deshalb seien hier einige Worte "uber die physikalischen 
Grundlagen angef"ugt (siehe auch \cite{Hu-et-al-1997}).

Im Mikrowellen-Hintergrund sehen wir weitgehend unverf"alscht die
Temperaturschwankungen auf der "`kosmischen Photosph"are"' bei der
Rotverschiebung $z\approx 1100$. Zu diesem Zeitpunkt kombinierten die 
bis dahin freien Elektronen und Atomkerne zu neutralen Atomen, und die 
W"armestrahlung entkoppelte von der Materie.
Die theoretische Analyse der Entwicklung von Dichte- und 
Temperaturschwankungen vor dieser "`Rekombinationszeit"' bietet 
bis auf die uns unbekannten Anfangsbedingungen keine prinzipiellen
Schwierigkeiten. 
Legt man f"ur letztere die gegenw"artig bevorzugten inflation"aren 
Modelle des sehr fr"uhen Universums zugrunde, so ergibt sich im
Leistungsspektrum der Temperaturanisotropien eine Sequenz von 
sogenannten "`akustischen Maxima"', deren Lage man recht einfach 
verstehen kann. Insbesondere entspricht das erste Maximum ungef"ahr 
der Ausdehnung des "`Schallhorizontes"' auf 
der kosmischen Photosph"are, d.h. der Distanz, welche eine Druckwelle 
bis zum Zeitpunkt der Rekombination durchlaufen kann. Der Winkel, unter 
welchem wir diese Ausdehnung sehen (er betr"agt etwa $1^{\circ}$),
h"angt im Wesentlichen nur von der r"aumlichen Kr"ummung ab. Dabei 
wirkt eine positive Raumkr"ummung auf Lichtstrahlen fokussierend (man 
denke sich als Analogon Lichtstrahlen auf einer Kugeloberfl"ache durch 
Gro"skreise gegeben) und daher wie bei einer Lupe vergr"o"sernd. 
Bei negativer Kr"ummung ist es umgekehrt. F"ur positive Kr"ummung 
(negatives $\Omega_K$) wird das Maximum also zu gr"o"seren 
Winkelabst"anden hin verschoben, f"ur negative Kr"ummung (positives
$\Omega_K$) zu kleineren. Einem flachen Universum entspricht 
eine Lage des ersten Maximums bei etwa $1^{\circ}$, bzw $\ell=200$ 
in der Entwicklung nach Kugelfunktionen.

% ABB 4 %%%%%%%%%%%%%%%%%%%%%%%%%%%%%%%%%%%%%%%%%%%%%%%%%%%%
%
\begin{figure}
\includegraphics[width=12cm]{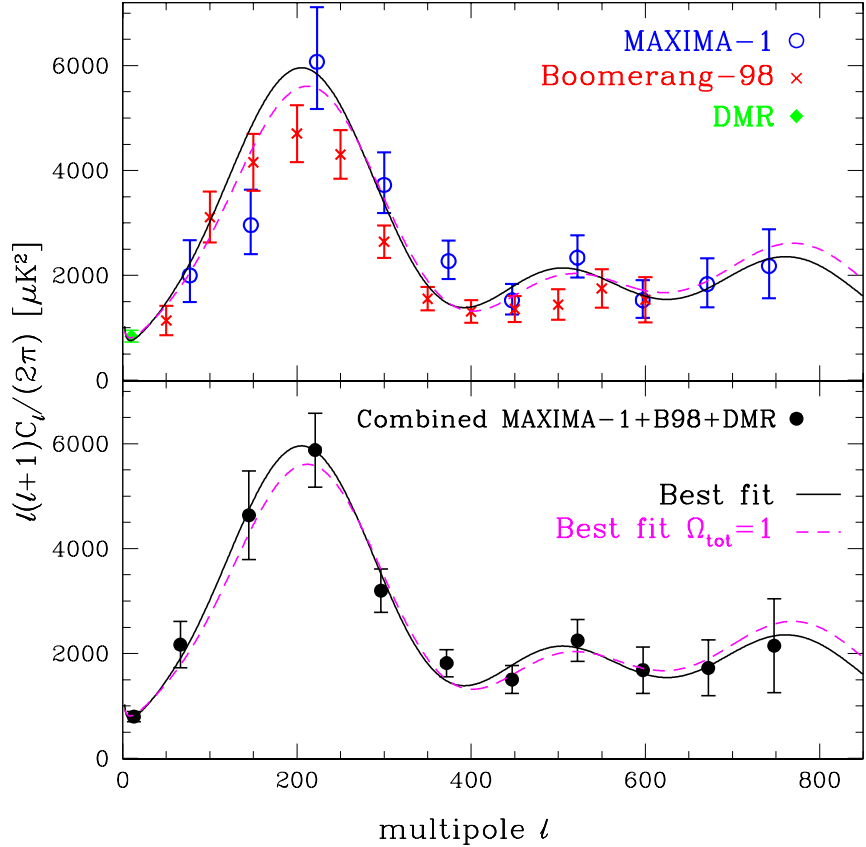}
\caption{\it
Leistungsspektrum der Anisotropien des Mikrowellenhintergrundes
der Experimente 
MAXIMA-1     (Fenster: $\ell=25-625$, Aufl"osung: $\delta\ell=50$) und 
BOOMERANG-98 (Fenster: $\ell=36-785$, Aufl"osung: $\delta\ell=75$).
$\ell$ bezeichnet wie "ublich die
"`Drehimpulsquantenzahl"' in der Multipolentwicklung.
Die Daten des Differential-Mikrowellen-Radiometers (DMR)
des Satelliten COBE \cite{Cobe} betreffen nur die kleinsten Werte 
von $\ell$. Im unteren Diagramm sind die Daten kombiniert.
Deutlich ist ein erstes Maximum bei $\ell\approx 200$,
entprechend einer Winkeldistanz von $\Delta\Theta=\pi/\ell\approx
1^{\circ}$ zu sehen. Das 95\% Konfidenzintervall f"ur 
$\Omega_{\rm tot}:=\Omega_{M}+\Omega_{\Lambda}$
betr"agt $\Omega_0=1.11\pm0.13$, ist also mit einem flachen Modell
$\Omega_{\rm tot}=1$ vetr"aglich. Die gestrichelten Kurven entsprechen
einer besten Anpassung eines flachen Modells mit den kosmologischen 
Parametern 
$(\Omega_B,\Omega_M,\Omega_{\Lambda},h)=(0.005,\,0.3,\,0.7,\,0.70)$.
Zusammen mit den Supernovae-Daten f"uhrt dies zur insgesamt besten
Anpassung. $\Omega_B$ bezeichnet den baryonischen Anteil von $\Omega_M$.
Das Diagramm ist \cite{Jaffe-et-al-2000} entnommen.}
\label{fig:Abb4}
\end{figure}       

Die Daten der neuesten Ballonexperimente 
"`Boomerang-98"'~\cite{Bernardis-et-al-2000}\cite{Lange-et-al-2000}
und  "`Maxima-1"'~\cite{Hanany-et-al-2000} f"uhren zu fast identischen 
%
% Abb.4 zum 1. mal genannt
%
Ergebnissen, wie die neuste gemeinsame Auswertung 
\cite{Jaffe-et-al-2000} zeigt, die in der oberen Abbildung~\ref{fig:Abb4} 
wiedergegeben ist. Insesondere ist ein erstes Maximum bei $\ell\approx 200$
"uberdeutlich. Richtige Pr"azisionsvermessungen wird der 
NASA-Satellit MAP (Microwave Anisotropy Probe \cite{MAP}) 
liefern, dessen Start f"ur den Herbst diesen Jahres vorgesehen ist.
Leider sind die Europ"aer viel langsamer: Der Start der ESA-Mission
PLANCK \cite{PLANCK} ist inzwischen auf 2007 verschoben worden.

\section*{Fazit und Ausblick}
Wir haben astronomische Beobachtungen besprochen, die es
jeweils gestatten, schmale aber langgestreckte Gebiete in 
der Ebene der zwei unabh"angigen kosmologischen Parameter 
$\Omega_M$ und $\Omega_{\Lambda}$ auszusondern. 
Zusammengenommen ergeben diese die ungef"ahren Werte 
\begin{equation}
(\Omega_M,\Omega_{\Lambda},\Omega_K,q_0)\approx
(1/3,2/3,0,-1/2),
\label{results}
\end{equation} 
wobei wir die abh"angigen Parameter $\Omega_K$ der Kr"ummung
und den Bremsparameter $q_0$ gleich mit aufgelistet haben.

\emph{Demnach befindet sich das gegenw"artige Universum in einem
Zustand beschleunigter Expansion, den es f"ur alle Zukunft
beibehalten wird (siehe Anhang~1). Es ist r"aumlich nahezu flach
und die Dichte an Materie, inklusive der dunklen Materie, macht
nur etwa 1/3 der kritischen Dichte aus. Den doppelten und damit
dominierenden Betrag an gravitativ wirksamer Energie- bzw.
Massendichte stellt der kosmologische Term.}

Hinsichtlich der oft gestellten Frage, ob wir in einem offenen
oder geschlossenen Universum leben, ist folgender mathematischer
Umstand zu beachten: Ein Universum positiv konstanter Kr"ummung ist
notwendig geschlossen, also auch von endlichem Volumen.
Ein flaches oder negativ gekr"ummtes Universum kann, muss aber
nicht offen, also von unendlichem Volumen sein. Hier werden die
Zusammenhangsverh"altnisse im Gro"sen \emph{nicht} durch die Geometrie
festgelegt. Die hier besprochenen Daten sind sowohl mit
verschwindender, als auch mit gen"ugend kleinen Kr"ummungen beider
Vorzeichen vertr"aglich.

Es ist nun sehr befriedigend, dass andere und physikalisch 
unabh"angige Methoden, wie z.B. Beobachtungen an reichen Galaxienhaufen, 
ebenfalls zu einem Wert $\Omega\approx 0,3\pm 0,1 (1\sigma)$ gelangen
(siehe \cite{Bahcall-et-al-1999}). Auch Massenbestimmungen mit Hilfe 
des Gravitationslinseneffektes st"utzen diesen Wert in sch"oner
Weise (siehe z.B.~\cite{Boehringer-et-al-1998}).
Dar"uberhinaus ist es m"oglich, den \emph{baryonischen Anteil}
$\Omega_B$ von $\Omega_M$ getrennt zu bestimmen. Dies geschieht 
einerseits mit Hilfe der primordialen H"aufigkeit der leichten 
Elemente, wie sie sich aus dem Urknallszenario der Nukleosynthese
ergibt, andererseits durch Bestimmung von Intensit"atsprofilen der
R"ontgenstrahlung des heissen Intrahaufengases. 
Dabei ergibt sich "ubereinstimmend, dass nur etwa 
$1/6$ der durch $\Omega_M$ repr"asentierten Materie baryonischer 
Natur ist. Die Natur des restlichen und "uberwiegenden Anteils der
Materie, die $\Omega_M$ ausmacht, ist uns bisher nicht bekannt.

\emph{
Zusammen mit dem noch unklaren Ursprung von $\Omega_{\Lambda}$
kann man also sagen, dass wir heute nur etwa $1/20$ der gravitativ
nachweisbaren Energie in Materie von uns heute bekannter Natur 
lokalisieren k"onnen. In Anbetracht dieser Tatsache erscheinen
gelegentliche Ank"undigungen des Endes der Physik 
etwas verfr"uht.}

Auf die k"urzlich vorgeschlagenen Versuche, das kosmische
Koinzidenzproblem zu l"osen, k"onnen wir hier nicht weiter 
eingehen. Die allgemeine Idee besteht darin, die kosmologische 
Konstante durch eine neue exotische Form von dynamischer Materie 
mit negativem Druck zu ersetzen. In Anlehnung an die urspr"ungliche 
Bedeutung dieses Wortes hat diese den Namen "`Quintessenz"' bekommen.
In konkreten Modellen wird sie durch ein Skalarfeld beschrieben, 
dessen Dynamik so eingerichtet ist, dass sich die Energiedichte 
des Feldes von selbst der Materiedichte angleicht, und dies 
weitgehend unabh"angig von den 
Anfangsbedingungen~\cite{Zlatev-et-al-1998}.
Die kosmologische Konstante wird also durch eine dynamische 
Gr"o"se ersetzt. Zuk"unftige genauere Beobachtungen werden es 
erm"oglichen, eine der beiden M"oglichkeiten zu verwerfen.

\section*{Anhang~1: 
          Dynamik der Fried\-mann-Lema\^{\i}tre Modelle
          mit kosmologischer Konstante}
In der Allgemeinen Relativit"atstheorie besteht ein kosmologisches Modell 
in der Angabe zweier Strukturen: 1)~einer 4-dimensionalen Raum-Zeit und
2)~einem zeitartigen Vektorfeld, das die (mittlere) Bewegung der 
Materie repr"asentiert. Die Klasse der Friedmann-Lema\^{\i}tre-Modelle 
ist definiert durch die Forderung, dass f"ur \emph{jeden} mit der Materie 
bewegten Beobachter Rotationssymmetrie herrscht. Daraus folgt, dass das 
4-dimensionale Linienelement in der Form 
\begin{equation}
ds^2=-dt^2+a^2(t)d\sigma^2
\label{Linienelement}
\end{equation}
geschrieben werden kann. (In diesem Anhang benutzen wir Einheiten, in
denen $c=1$ ist.) Darin ist $t$ die kosmische Zeit, die durch
die Eigenzeit der Materie definiert wird und $d\sigma^2$ ist ein 
3-dimensionales Linienelement einer r"aumlichen Geometrie von 
normierter konstanter Kr"ummung $\pm 1$ oder $0$. Die einzige
dynamische  Variable ist der globale r"aumliche Skalenfaktor $a(t)$, 
der angibt, wie sich physikalische Abst"ande zwischen Raumpunkten 
gleicher kosmischer Zeit $t$ mit dieser Zeit "andern.

Setzt man (\ref{Linienelement}) in die Einstein-Gleichung mit 
kosmologischer Konstante $\Lambda$ ein, wobei die Materie durch eine 
Massendichte $\rho_M$ und einen Druck $p_M$ repr"asentiert wird,
so erh"alt man die zwei unabh"angigen Gleichungen:
\begin{eqnarray}
\left(\frac{\dot a}{a}\right)^2
&=&
\frac{8\pi G}{3}\rho_M+\frac{\Lambda}{3}-\frac{k}{a^2}
\label{Friedmann-1}\,,\\
\ddot a
&=&
-\frac{4\pi G}{3}(\rho_M+3p_M)\,a +\frac{\Lambda}{3}a
\label{Friedmann-2}\,.
\end{eqnarray}
Diese implizieren eine Art Energieerhaltungsgleichung:
\begin{equation}
{\dot\rho}_M=-3\frac{\dot a}{a}(\rho_M+p_M)\,.
\label{Friedmann-3}
\end{equation} 
F"ur $\dot a\not =0$ ziehen je zwei dieser Gleichungen 
die dritte nach sich und repr"asentieren damit die gesamte 
Information der Einsteinschen Feldgleichungen. Ist $p_M\not =0$, 
so muss  zus"atzlich noch eine Zustandsgleichung $p_M(\rho_M)$
hinzukommen.

Zum gegenw"artigen Zeitpunkt der Entwicklung des Universums, 
der mit $t_0$ bezeichnet sei, dominiert aber die nichtrelativistische 
Materie, so dass der Druck $p_M$ vernachl"assigbar ist. 
Gleichung (\ref{Friedmann-3}) impliziert dann sofort die zeitliche 
Konstanz von $\rho_Ma^3$. Der Hubble-Parameter ist definiert durch 
$H_0={\dot a}_0/a_0$, wobei hier und im Folgenden der Index $0$ 
die Auswertung zur Zeit $t_0$ bezeichnet.  Betrachtet man  
(\ref{Friedmann-1}) zum Zeitpunkt $t_0$ und dividiert die Gleichung  
durch $H_0^2$, so folgt mit~(\ref{def-omega})
\begin{equation}
\Omega_M+\Omega_{\Lambda}+\Omega_K=1\,,
\label{triangle}
\end{equation}
wobei $\Omega_K:=-k/a_0^2H_0^2$. 

Der \emph{Bremsparameter} $q_0:=-{\ddot a}_0/a_0H_0^2$ bestimmt, 
ob die gegenw"artige Expansion verz"ogert ($q_0>0$) oder 
beschleunigt ($q_0<0$) verl"auft. Addiert man
$2\times\hbox{(\ref{Friedmann-2})}$
zu (\ref{Friedmann-1}) und benutzt (\ref{triangle}) so erh"alt man 
\begin{equation}
q_0=\shalf\Omega_M-\Omega_{\Lambda}\,.
\label{Bremsparameter}
\end{equation}

Zum Verst"andnis der Dynamik bei verschwindendem Materiedruck 
ist es zweckm"a"sig, Gleichung (\ref{Friedmann-1}) in folgender
dimensionsloser Form zu schreiben: Man setzt
$\tau:=H_0t$ und $x(\tau):=a(t)/a(t_0)$; 
mit $\rho_Ma^3=\hbox{konst}$. erh"alt man
\begin{eqnarray}
&&
\left(\frac{dx}{d\tau}\right)^2+U(x)=\Omega_K\,,
\label{Teilchen}\\
\hbox{mit}
&&
U(x)=-\Omega_M\,x^{-1}-\Omega_{\Lambda}x^{2}\,,
\label{Potential}
\end{eqnarray}
was (modulo eines unwichtigen globalen Faktors $\shalf$) als 
Energiesatz einer eindimensionalen Bewegung eines Teilchens im 
Potential $U(x)$ und Gesamtenergie $\Omega_K$ gem"a"s der 
Newtonschen Mechanik interpretiert werden kann. F"ur
$\Omega_{\Lambda}>0$ hat $U$ genau ein Maximum. Rollt das
Universum "uber dieses hinweg, so setzt eine unendliche Zeit
beschleunigter Expansion ein, in der wir uns gegenw"artig zu
befinden scheinen. Analog lassen sich mit (\ref{Potential})
%
% Abb 1 zum 5. mal genannt 
%
die in Abbildung~\ref{fig:Abb1} eingezeichneten Grenzkurven f"ur die 
Dynamik nachvollziehen. Beachte insbesondere, dass $q_0=0$
gem"a"s (\ref{Bremsparameter}) der Geraden 
$\Omega_{\Lambda}=\frac{1}{2}\Omega_M$ entspricht. 

Einsteins statisches Universum entspricht der L"osung  
$\Lambda=a^{-2}=4\pi G\rho_M$ und $k=1$ von 
(\ref{Friedmann-1},\ref{Friedmann-2}), die am Maximum des 
Potentials (\ref{Potential}) verweilt und deshalb instabil 
ist. De~Sitters Universum entspricht der L"osung f"ur 
$\rho_M=p_M=0$ und $k=1$, die, wie man aus (\ref{Friedmann-2}) 
sofort sieht, f"ur gro"se $t$ exponentiell expandiert.

\section*{Anhang~2: Vakuumfluktuationen in der Quantenfeldtheorie}
In diesem Anhang benutzen wir Einheiten, in denen $c=1$ und 
$\hbar=1$ ist.

In der Quanten(feld)theorie f"uhren die notwendigerweise
auch im Vakuumzustand existenten \emph{Schwankungen} zu einer
nicht verschwindenden Energie. Dies sei zun"achst am harmonischen
Oszillator erl"autert. 

Dessen Energieoperator als Funktion des Ortsoperators $q$ und
Impulsoperators $p$ ist bekanntlich gegeben durch 
\begin{equation}
H=\shalf p^2/m+\shalf m\omega^2 q^2\,,
\label{Harm-Oszillator}
\end{equation}
mit $m$ und $\omega$ als seine Masse und Kreisfrequenz.

Die kanonischen Vertauschungsrelationen $[q,p]=i$
bedingen nun eine nicht verschwindende Energie f"ur den 
Grundzustand. Da n"amlich die Erwartungswerte von $q$ und 
$p$ verschwinden, sind die Erwartungswerte ihrer Quadrate
gleich den Schwankungsquadraten. Die Schwankungen selbst gen"ugen 
aber der Unsch"arferelation 
$\Delta q\cdot\Delta p\geq\shalf$, die somit verhindert, 
dass die Erwartungswerte f"ur potentielle und kinetische Energie 
gleichzeitig beliebig klein werden k"onnen. Als Kompromiss ergibt 
sich eine minimale Gesamtenergie (Grundzustand) von
$\shalf\omega$, zu der beide Teilenergien gleich beitragen.

Das gleiche Ph"anomen existiert nun bei quantisierten Feldern.
Nehmen wir z.B. das freie elektromagnetische Feld: Hier ist
die Energiedichte gegeben durch 
\begin{equation}
\rho=\shalf 
({\vec E}^2(\vec x)+
{\vec B}^2(\vec x))\,.    
\label{E-M-Feld}
\end{equation}
Der Ortsvariablen $q$ entspricht hier das Vektorpotential $\vec A(\vec x)$,
wobei $\vec B=\vec\nabla\times\vec A$, der Impulsvariablen das Negative
des elektrischen Feldes $-\vec E(\vec x)$. Diese erf"ullen zu gleichen 
Zeiten wieder die kanonischen Vertauschungsrelationen, aus denen 
man durch einfache Differentiation den Kommutator zwischen 
elektrischem und magnetischem Feld zu gleichen Zeiten 
erh"alt (Jordan und Pauli, 1928):
\begin{equation}
\left[E_i(\vec x),B_{jk}(\vec y)\right]
= i\left(\delta_{ij}\partial_k 
-        \delta_{ik}\partial_j\right)\,
\delta^{(3)}(\vec x-\vec y)\,,
\label{Pauli-Jordan}
\end{equation}
wobei $B_{12}:=B_3$ und zyklisch. 
Im Grundzustand verschwinden wieder die Erwartungswerte der 
Feldst"arken, so dass seine Energiedichte wieder eine Summe von 
Schwankungsquadraten des elektrischen und magnetischen Anteils 
ist, die wegen (\ref{Pauli-Jordan}) nicht gleichzeitig klein
gemacht werden k"onnen. Der energetische Grundzustand, genannt
\emph{Vakuum}, ergibt sich wieder aus einem Kompromiss minimaler
Unsch"arfen in beiden Feldern, wobei beide Terme in
(\ref{E-M-Feld}) gleich beitragen. Gegen"uber der Quantenmechanik 
besteht hier jedoch noch die mathematische Notwendigkeit der 
Regularisierung, ohne die Ausdr"ucke wie $\vec E^2(\vec x)$
(d.h. Quadrate von Feldoperatoren am gleichen Punkt) keinen 
mathematischen Sinn haben. Dazu verschmiert man die 
Feldoperatoren am Punkt $\vec x$ mit Testfunktionen $f$, d.h. 
man bildet
\begin{equation}
\vec E_f(\vec x):=\int d^3y\,{\vec E(\vec x+\vec y)f(\vec y)}
\label{verschmiert}
\end{equation}
und berechnet den Vakuumerwartungswert des Quadrats. Es ergibt 
sich
\begin{equation}
\langle{\vec E}_f^2(\vec x)\rangle_{\rm vac} = 
2\int\frac{d^3k}{(2\pi)^3}\frac{\omega}{2}
\vert\hat f(\vec k)\vert^2\,,
\label{E-quadrat}
\end{equation}
wo $\hat f$ die Fouriertransformierte von $f$ im $\vec k$-Raum ist
und $\omega=\vert\vec k\vert$. 

Die r"aumlich Energiedichte des Vakuums, die wegen der
Translationsinvarianz des Vakuums konstant sein muss, ist 
einfach durch (\ref{E-quadrat}) gegeben, da elektrischer und 
magnetischer Anteil in (\ref{E-M-Feld}) gleich beitragen. 
Dieser Ausdruck divergiert, falls $\hat f$ nicht 
gen"ugend schnell abf"allt (Ultraviolett-Divergenz). W"ahlt man 
$\hat f\equiv 1$ f"ur Frequenzen kleiner einer Maximalfrequenz 
$K$ und $\equiv 0$ f"ur Frequenzen dar"uber, so ergibt sich 
die $K$-abh"angige Vakuumenergiedichte zu
\begin{equation}
\rho_{\rm vac}=\frac{K^4}{8\pi^2}\,.
\label{rho-vac}
\end{equation}

Ohne Gravitation k"ummern wir uns nicht um die Energiedichte
des Vakuums, da dann nur Energie\emph{differenzen} beobachtbar 
sind. Aber auch dann k"onnen Quantenfluktuationen des Vakuums
wichtig sein, wie z.B. beim Casimir-Effekt deutlich wird.
Dort werden durch die Anwesenheit leitender Platten 
unterschiedliche Randbedingungen an das elektromagnetische Feld 
gestellt und die verschiedenen Konfigurationen verglichen. 
F"ur die beobachtbaren Gr"o"sen, wie z.B. die Anziehungskraft der
erw"ahnten Platten, ergeben sich dabei stets Ausdr"ucke, die im 
Limes $K\mapsto\infty$ endlich bleiben und nun auch in
Pr"azisionsexperimenten verifiziert wurden. Siehe dazu 
\cite{Kardar-Golestanian-1999} und die dort angegebenen 
Referenzen.

Die \emph{gravitative} Wirkung der Vakuumfluktuationen muss nach 
den Einstein-Gleichungen "uber den Erwartungswert ihres 
Energie-Impuls-Tensors laufen, der aus Invarianzgr"unden gerade 
die Form eines 
kosmologischen Terms mit $\Lambda_{\rm vac}=8\pi G\rho_{\rm vac}$ 
hat. Zusammen mit einem eventuell a priori vorhandenem $\Lambda_0$,
das keiner zus"atzlichen Energieform entspr"ache sondern als neue
Naturkonstante den Vakuum-Einsteingleichungen zugeh"orig w"are,
ergibt sich dann ein effektives $\Lambda=\Lambda_0+\Lambda_{\rm vac}$
f"ur das dann aus den bereits genannten Gr"unden $\rho_{\Lambda}$
wiederum nicht wesentlich gr"o"ser als $<\rho_{\rm krit}$ sein kann. 
F"ur eine Abschneideskala $K$ in der Gr"o"senordnung $K\approx 0,1\,GeV$
(QCD-Phasen"ubergang) m"ussten $\Lambda_0$ und 
$\Lambda_{\rm vac}$ auf "uber 40 Stellen genau entgegengesetzt gleich
sein; bei $K\approx 100\,GeV$ (Fermi-Skala)  m"usste diese
Genauigkeit sogar auf "uber 52 Stellen genau gelten. Das Problem der
kosmologischen Konstanten stellt sich dann in Form einer unerkl"arten 
extremen Feinabstimmung.

\end{document}